\documentclass[a4paper,11pt]{article}

\usepackage{tikz}
\usetikzlibrary{backgrounds}
\usepackage{caption}

\usepackage{amsmath,amssymb}    
\usepackage{color}
\usepackage{graphicx}
\usepackage{subfigure}
\usepackage{cite}               
\usepackage{hyperref}           

\numberwithin{equation}{section}   

\def \be {\begin{equation}}
\def \ee {\end{equation}}
\def \ba {\begin{array}}
\def \ea {\end{array}}
\def \bea{\begin{eqnarray}}
\def \eea{\end{eqnarray}}

\def \d {\delta}

\def \p {\partial}

\def \nn {\nonumber}

\def \hs {\hspace}

\def \lag {\langle}
\def \rag {\rangle}

\def \mO {\mathcal O}

\begin{document}
\title{Note on Bulk Reconstruction in AdS$_3$/WCFT$_2$}

\author{Yu-han Lin$^{1}$  and Bin Chen$^{1,2,3}$ \thanks{bchen01@pku.edu.cn,  1301110085@pku.edu.cn}}
\maketitle
\begin{center}
\textsl{$^{1}$Department of Physics and State Key Laboratory of Nuclear
Physics and Technology, Peking University, No.5 Yiheyuan Rd, Beijing
100871, P.R. China}\\
\textsl{$^{1}$Center for High Energy Physics, Peking University,
No.5 Yiheyuan Rd, \\Beijing 100871, P. R. China}\\
\textsl{$^{3}$Collaborative Innovation Center of Quantum Matter,
No.5 Yiheyuan Rd, \\Beijing 100871, P. R. China}\\
\par\end{center}

\date{}

\maketitle

\vspace{-10mm}

\vspace{8mm}

\begin{abstract}
The bulk reconstructions in AdS/CFT and its cousins are essential to understand the holographic nature of quantum gravity. In this work, we try to study 
 the bulk reconstruction in the AdS$_3$/WCFT$_2$ correspondence. After deriving the bulk-boundary propagator, which is different from the usual one in AdS/CFT, we define the bulk proto-scalar field by using the Virasoro-Kac-Moody symmetry in two different ways. One is to impose the bulk primary conditions on the field and construct the field algebraically. The other is to use the bulk-boundary vacuum OPE block, which can be read by applying the diffeomorphism preserving the CSS boundary conditions. Two approaches lead to consistent picture.  
 

\end{abstract}
\baselineskip 18pt

\thispagestyle{empty}
\newpage

\tableofcontents

\section{Introduction}

The AdS/CFT correspondence states that the  quantum gravity in Anti-de Sitter (AdS) spacetime is dual to a conformal field theory at asymptotical AdS boundary \cite{Maldacena:1997re}.  It has been intensely investigated from various points of view since its proposal.   One central question is to understand how the holography works in the correspondence. This inspired  people to study the holographic duality to the spacetime beyond the AdS or the quantum field theory without full conformal symmetry. Such kind study has resulted in the setup of other correspondences beyond AdS/CFT, including dS/CFT\cite{Strominger:2001pn}, Kerr/CFT\cite{Guica:2008mu}, Warped AdS/CFT\cite{Anninos:2008fx}, AdS/WCFT\cite{Compere:2013bya} etc.. 

It turns out that  the choice of asymptotic boundary conditions and the corresponding asymptotic symmetry group(ASG) plays an essential role in setting up the correspondences. In particular, for the AdS$_3$ gravity, it has been found that besides the well-known Brown-Henneaux boundary conditions\cite{Brown:1986nw} which leads to the AdS$_3$/CFT$_2$ correspondence\cite{Strominger:1997eq}, there exists other sets of consistent boundary conditions \cite{Compere:2013bya,Grumiller:2016pqb}. One set of them is the Comp\`ere-Song-Strominger (CSS) boundary conditions, whose ASG is generated by  an Virasoro-Kac-Moody algebra. It was proposed that the AdS$_3$ gravity with CSS boundary conditions should be dual to a warped CFT, which has the same Virasoro-Kac-Moody symmetry\cite{Compere:2013bya}.  
This so-called AdS$_3$/WCFT$_2$ correspondence\footnote{There exists another correspondence which states that the gravity in a warped AdS$_3$ may be dual to a warped CFT\cite{Castro:2015csg,Song:2016gtd}. This  correspondence relied on the fact that the isometry group of warped AdS$_3$ spacetime is the same as the warped conformal symmetry. However, the warped spacetime can only appear in the topological gravity theory\cite{Anninos:2008fx}. In this work, we focus on the AdS$_3$/WCFT correspondence. } has been studied from several points of view, including the microscopic entropy of black holes\cite{Detournay:2012pc}, the entanglement entropy\cite{Castro:2015csg,Song:2016gtd,Apolo:2018eky}, the 1-loop partition function\cite{Castro:2017mfj} and very recently the R\'enyi mutual information of two disjoint intervals\cite{Chen:2019xpb}. In this work, we would like to study the bulk reconstruction in the AdS$_3$/WCFT$_2$ correspondence.

The bulk reconstruction is one of most important questions in the AdS/CFT correspondence.  As a concrete realization of holographic principle in quantum gravity, the AdS/CFT correspondence provides a way to probe bulk physics from boundary data.  In a narrow sense, the bulk reconstruction is an attempt to express the bulk field in terms of boundary local operators.  It allows us to understand bulk processed using CFT techniques\cite{Banks:1998dd}. There are several attempts to investigate the bulk reconstruction, for a nice review see \cite{Harlow:2018fse}. The HKLL global reconstruction \cite{Hamilton:2005ju,Hamilton:2006az} is the earliest and mostly studied one. In 2d CFT, the exsitence of local conformal algebra allows local bulk reconstruction.  Recently an exact AdS$ _3 $ proto-field \cite{Anand:2017dav,Chen:2017dnl} has been proposed to capture the non-perturbative information of gravitational background. The study relied heavily on the symmetry of the theory. From the symmetry considerations, a bulk primary condition is imposed on and can be used to solve the proto-field exactly. 

In this work, we study the bulk reconstruction via the proto-field. As the first step, we need to study the global reconstruction in the AdS/WCFT correspondence. We propose a bulk-boundary propagator for the boundary primary operator. Similar to the AdS/CFT case, we start by solving the Klein-Gordon equation for the bulk-boundary propagator. But now as the isometry group is broken by the choice of the boundary condition, we cannot turn on the delta-function source to the Klein-Gordon equation. Instead we have to require that the bulk-boundary propagator matches the two point function of WCFT as the bulk point approaches the boundary. With the bulk-boundary propagator, we compute the bulk 3-point Witten diagrams and find agreement with the  3-point functions in WCFT. Using  the bulk-boundary propagator obtained, we apply the HKLL proposal to construct the bulk field and express it  in terms of the the boundary primary and its global descendants, similar to the AdS/CFT case. And using the bulk field we get, we calculate the bulk-bulk propagator. Furthermore, we show how to  define a  proto-field by imposing  bulk primary conditions. We solve  the proto-field exactly. We have also considered the correlator of two bulk proto-fields, but a compact result has not been found. We take another way to describe the bulk proto-field by using the bulk-boundary OPE block. Using a diffeomorphism between AdS spaces compatible with the  CSS boundary conditions and the vacuum AdS space, we obtain the bulk-boundary vacuum OPE block for AdS/WCFT correspondence.


The rest of the paper goes as follows. In section 2 we give a brief review on the AdS/WCFT correspondence. In section 3, we study the bulk-boundary propagator in the AdS/WCFT correspondence. We discuss the  requirement for the bulk-boundary propagator and find the propagator. And then we construct the bulk field following the HKLL proposal.  In section 4 we study the local reconstruction  in AdS/WCFT. We impose the bulk primary condition on  the proto-scalar field, which can be solved level by level. We also discuss the properties of the correlator between two bulk level $n$ proto-fields. Moreover we investigate the bulk proto-field by studying its  OPE block with the boundary primary. We focus on the bulk-boundary vacuum OPE  block, which can be read from the diffeomorphism preserving the CSS boundary conditions and is also most related to the bulk gravitational physics.

\section{The AdS/WCFT Correspondence}

This section is a brief review on the AdS/WCFT correspondence. Both the bulk and boundary perspectives are introduced. Some basic concepts and properties are presented for later use.


A two dimensional translation-invariant theory with a chiral scaling symmetry could be extended to a local algebra\cite{Hofman:2011zj}. There are two minimal options for this algebra. One is the usual conformal algebras with two copies of Virasoro algebras, and the other is warped conformal algebra with a Virasoro algebra plus a U(1) Kac-Moody algebra. The latter choice is what we called warped conformal field theory\cite{Detournay:2012pc}. The local symmetry of a WCFT is  of the form
\begin{equation}x^+ \to f(x^+),\hs{3ex}x^- \to x^-+g(x^+).\end{equation}
When the theory is put on a cylinder, the conserved charges can be written in terms of the Fourier modes. And the symmetries of WCFT is generated by a copy of Virasoro algebra with central charge $c$ and a copy of Kac-Moody algebra with level $k$. The commutation relations of the generators are
\begin{equation}\begin{aligned}
&[L_n,L_m]=(n-m)L_{n+m}+\frac{c}{12}n(n^2-1)\delta_{n,-m} ,\\
&[L_n,P_m]=-mP_{m+n} ,\\
&[P_n,P_m]=\frac{k}{2}n\delta_{n,-m},
\end{aligned}\end{equation}
where the Kac-Moody generators transform canonically under the action of the Virasoro generators. After subtracting the Sugawara construction of the U(1) current from the energy-momentum tensor, the resulting Virasoro algebra and Kac-Moody algebra commute with each other. With respect to the symmetries, the states of WCFT can be described by the representations of the algebra. The primary states are defined to obey the relations 
\begin{equation}\begin{aligned}
&L_n|\Delta,Q>=0,\ P_n|\Delta,Q>=0, \quad \forall n>0\\
&L_0|\Delta,Q>=\Delta|\Delta,Q>,\ P_0|\Delta,Q>=Q|\Delta,Q>.
\end{aligned}\end{equation}
Namely, the primary states are characterized by the scaling dimension $\Delta$ and the charge $Q$. And the states in the highest weight representation is the descendants of the primary state by acting on the creation operators. In this paper only the primary states are of our concern\footnote{For other aspects on the warped CFT, see \cite{Hofman:2014loa,Castro:2015uaa,Apolo:2018oqv}. }. 

 Three-dimensional  AdS gravity should be defined properly with the asymptotical boundary conditions. 
  When imposing the Brown-Henneaux(BH) boundary conditions\cite{Brown:1986nw} on AdS$_3$ gravity, the asymptotic symmetry group on the boundary are generated by two copies of the Virasoro algebra with central charges $c_L=c_R=\frac{3l}{2G}$. This suggests that the AdS$_3$ gravity with BH boundary conditions could be dual to a 2D CFT. The corresponding dual conformal field theory is called holographic CFT, which should have a large central charge and a sparse spectrum of light states\cite{Hartman:2014oaa}. 
  
  The AdS$_3$ gravity can be defined with other consistent boundary conditions. Under the CSS boundary conditions, the ASG is generated by the Virasoro-Kac-Moody algebra\cite{Compere:2013bya},  but not in the canonical form
\begin{equation}\begin{aligned}
&[L_n,L_m]=(n-m)L_{n+m}+\frac{c}{12}n(n^2-1)\delta_{n,-m} \\
&[L_n,P_m]=-mP_{m+n}+mP_0\delta_{n,-m} \\
&[P_n,P_m]=-2nP_0\delta_{n,-m}.
\end{aligned}\end{equation}
The central charge takes the same value  as in the Brown-Henneaux case
\be
 c=3l/2G.
 \ee
  The asymptotic symmetry algebra and the canonical WCFT algebra are actually related by a charge redefinition.
\begin{equation}L_n\to L_n+2P_0P_n-P_0^2\delta_n,\quad P_n\to 2P_0P_n-P_0^2\delta_n.\end{equation}
This suggests that the AdS$_3$ gravity with the CSS boundary condition could be dual to a WCFT. 
 The  corresponding holographic WCFT is expected to have a large central charge, a negative Kac level and a sparse spectrum of light states. 
 The regularized on-shell bulk gravity action is now a chiral Liouville gravity\cite{Compere:2013aya}. 
  
 
The CSS boundary conditions are chiral boundary conditions with one component of the boundary metric being promoted to a dynamical variable. The most general solution to the Einstein's equations compatible with these boundary conditions can be written in the Fefferman-Graham gauge as follows
\begin{equation}\begin{aligned}ds^2=&\frac{dy^2+dz(d\bar{z}-B'(z)dz)}{y^2}-\frac{6}{c}L(z)dz^2-\frac{6}{c}\bar{L}(d\bar{z}-B'(z)dz)^2\\&
+\frac{36}{c^2}L(z)\bar{L}y^2dz(d\bar{z}-B'(z)dz),\end{aligned}\label{BanadosWAdS}
\end{equation}
where $\bar{L}$ is a constant and $B'(z)\equiv \p_z B$ are only the functions of $z$. 
When $B'=0$, this metric can be viewed as a Banados solution  with constant anti-holomorphic stress tensor.

\section{Global bulk reconstruction in AdS/WCFT} 

In this section, we study the global bulk reconstruction in AdS/WCFT, which based mainly on the global symmetry of the correspondence. As discussed above, the global symmetry is now simply $SL(2,R)\times U(1)$ symmetry. We would like to follow the line of HKLL global reconstruction, but due to the isometry of the AdS$_3$ spacetime is broken by the asymptotical boundary conditions, we need to pay more attention to the boundary conditions. 

\subsection{Bulk-boundary Propagator in AdS/WCFT}

We start from the construction of bulk-boundary propagator in the AdS/WCFT correspondence. We first study the bulk-boundary propagator for a WCFT primary in the vacuum. In the following, we use the Poincare coordinates to describe the AdS$ _3 $ spacetime. The coordinates are $ (y,z,\bar{z}) $ with the vacuum AdS metric being
\be
 ds^2=\frac{dy^2+dzd\bar{z}}{y^2}. 
 \ee
  The asymptotic boundary corresponds to $y \to 0$. 

\par From the extrapolate dictionary, the bulk field should be related to the boundary primary as it approaches the boundary
\be
\phi(y,z)|_{y \to 0}=y^{2h} \mO(z). 
\ee
 The bulk-boundary propagator should recover a boundary WCFT correlator as the bulk point approaches the boundary. The correlator of two primaries in a  WCFT can be determined by the global symmetries\cite{Song:2017czq}
\be
 G(z_1, z_2)=\lag \mO(z_1)\mO(z_2)\rag=e^{iQ(\bar{z}_1-\bar{z}_2)}\frac{1}{(z_1-z_2)^{2h}}, 
 \ee
 where the two primaries are required to have the opposite charges and the same scaling dimension $h$.  Without loss of generality, the bulk field  could be put at $ (y,z,\bar{z}) $ and the boundary primary  at the origin $(0,0)$. Then the bulk-boundary propagator $ K(y,z,\bar{z};0,0) $ has to satisfy the requirement
 \begin{equation}K(y,z,\bar{z};0,0)|_{y\to0}=y^{2h}G(z,0)=y^{2h}z^{-2h}e^{iQ\bar{z}}.\end{equation}
Another basic requirement for the bulk-boundary propagator is to satisfy the Klein-Gordon equation
 \begin{equation}
 (\Box-m^2)K(y,z,\bar{z};0,0)=0.
 \end{equation}
As the AdS$_3$ spacetime has the isometry group $SL(2,R)\times SL(2,R)$, the solutions to the above equation may respect the symmetry without imposing boundary conditions. With the Brown-Henneaux boundary condition, the isometry group is kept intact such that the boundary CFT has the full conformal symmetry. This is also reflected in the fact that in the Brown-Henneaux case, the righthand side of the above equation could have a delta-function source. 
On the contrary, the CSS boundary conditions obviously break the full isometry group to $SL(2,R) \times U(1)$. To solve the Klein-Gordon equation, we need to take the boundary conditions appropriately. For a general primary field, it is not clear how to impose the boundary condition. Due to the plane wave form of the $ \bar{z} $ dependence, it is not possible to have a delta-function source. So we only require the Klein-Gordon equation to be satisfied when the bulk and boundary point are separated.


The above two requirements cannot determine the bulk-boundary propagator completely. To  solve the problem, we would like to make the  ansatz   that the dependence of the bulk-boundary propagator on $ \bar{z} $ is simply $ e^{iQ\bar{z}} $, considering the form of the WCFT correlators. A natural explanation is that we consider only the Kac-Moody U(1) primary, so to transform correctly under the level zero U(1) generator, this form of $ \bar{z} $ dependence is needed. With this ansatz,  we can expand the propagator in a power series
\begin{equation}
K(y,z,\bar{z};0,0)=y^{2h}z^{-2h}e^{iQ\bar{z}}\sum_{n=0}^{\infty}C_n(\frac{y^2}{z})^n,
\end{equation}
where $C_n$'s are the coefficients to be determined. 
Substituting the expression into the Klein-Gordon equation, we can solve the equation order by order and get the relations for the coefficients. From the zeroth order, we obtain the mass equation which is the same as usual AdS/CFT correspondence
\be
m^2=2h(2h-2). 
\ee
And from the orders above zero, we obtain a recursive relation for the coefficients
\begin{equation}C_n=\frac{iQ}{n}C_{n-1}. \end{equation}
They are the expansion coefficients of an exponential function. Finally we obtain the bulk-boundary propagator 
\begin{equation}
K(y,z,\bar{z};0,0)=y^{2h}z^{-2h}e^{iQ\bar{z}}e^{\frac{iQy^2}{z}}=y^{2h}z^{-2h}e^{iQ\frac{z\bar{z}+y^2}{z}}.
\end{equation}
We can compare it to the usual AdS$ _3 $ bulk-boundary propagator 
\be
K(y,z,\bar{z};0,0)=(\frac{y}{y^2+z\bar{z}})^{2h} .\ee
 The AdS/CFT bulk-boundary propagator only depend on the geodesic distance between the bulk and the boundary points. For the AdS/WCFT bulk-boundary propagator, this is not the case.  

In the usual AdS/CFT correspondence, one may reconstruct a bulk field in terms of a boundary primary and its global descendants by requiring it  reproduce the right bulk-boundary propagator.
  It is the global bulk field in the HKLL reconstruction\cite{Hamilton:2005ju,Hamilton:2006az}
\begin{equation}
\phi(y,z,\bar{z})=\sum_{n=0}^{\infty}y^{2h+2n}\frac{(-1)^n}{n!(2h)_n}L_{-1}^n\bar{L}_{-1}^n\mO,
\end{equation}
where $\mO$ is the boundary primary operator and $ \phi $ is the bulk field. 

In the AdS/WCFT correspondence, we can have the similar reconstruction. Comparing the structure of Virasoro-Kac-Moody algebra with two copies of Virasoro algebra , we see that the $U(1)$ generator in the Kac-Moody sector replaces the Virasoro  generator in anti-holomorphic sector. This leads us to 
replace the global anti-holomorphic Virasoro generator in the AdS/CFT bulk field with the global U(1) generator to get the AdS/WCFT bulk field in the HKLL proposal. This gives us 
\begin{equation}
\phi(y,z,\bar{z})=\sum_{n=0}^{\infty}y^{2h+2n}\frac{(-iQ)^n}{n!(2h)_n}L_{-1}^n\mO. 
\end{equation}
One may check that this bulk field indeed gives the right bulk-boundary propagator.

\par Using the expression for a global bulk field, we can calculate the bulk-bulk propagator between two points $ (y_1,z_1,\bar{z}_1) $ and $ (y_2,z_2,\bar{z}_2) $. It is done by first calculating the correlator of the two primaries, then acting on the Virasoro operators. As the holomorphic Virasoro generators can be expressed as holomorphic derivatives,  the bulk-bulk propagator can be computed
\begin{equation}\begin{aligned}
\langle \phi_1\phi_2\rangle&=y_1^{2h}y_2^{2h}e^{iQ\bar{z}_{12}}\sum_{n_1=0,n_2=0}^{\infty}\frac{(-iQy_1^2)^{n_1}(-iQy_2^2)^{n_2}{\partial_{z_1}}^{n_1}{\partial_{z_2}}^{n_2}(\frac{1}{z_{12}^{2h}})}{n_1!(2h)_{n_1}n_2!(2h)_{n_2}}\\
 &=\frac{y_1^{2h}y_2^{2h}}{z_{12}^{2h}}e^{iQ\bar{z}_{12}}\sum_{n_1=0,n_2=0}^{\infty}\frac{(2h)_{n_1+n_2}}{n_1!n_2!(2h)_{n_1}(2h)_{n_2}}(\frac{iQy_1^2}{z_{12}})^{n_1}(\frac{iQy_2^2}{z_{12}})^{n_2}. 
\end{aligned}\end{equation}
It is a bi-variate hypergeometric function and cannot be simplified to a closed form. It satisfies some nice properties. Firstly the bulk-bulk propagator simplifies to the bulk-boundary propagator as one point approaches the boundary. Secondly we can check that the bulk-bulk propagator satisfies the Klein-Gordon equation.

\subsection{ Three-point Witten diagrams}

As a consistent check, we compute the bulk three-point Witten diagram, which should capture  the three-point correlation functions in holographic WCFT. 
From global symmetry consideration, the most general form of 3-point functions of primary operators in WCFT is\cite{Song:2017czq}
\begin{equation}\begin{aligned}
&\langle O(z_1,\bar{z}_1)O(z_2,\bar{z}_2)O(z_3,\bar{z}_3\rangle \\
&=C\delta_{Q_1+Q_2+Q_3,0}\frac{e^{\frac{1}{3}(Q_1-Q_2)\bar{z}_{12}}e^{\frac{1}{3}(Q_2-Q_3)\bar{z}_{23}}e^{\frac{1}{3}(Q_3-Q_1)\bar{z}_{31}}}{z_{12}^{h_1+h_2-h_3}z_{23}^{h_2+h_3-h_1}z_{31}^{h_3+h_1-h_2}}, 
\end{aligned}\end{equation}
where $C$ is a constant that depends on the theory. The form of the 3-point function should be reproduced from the bulk computation of the Witten diagrams.

The Witten diagram calculation for the 3-point function of a primary is to integrate over a bulk point the product of three bulk-boundary propagators between it and the boundary points. Now we have to use the  bulk-boundary propagators in AdS/WCFT. The 3-point vertex depends on the particular theory and  is not important if we only care the form of 3-point function. The expression for the Witten diagram is
\begin{equation}\begin{aligned}
G&\propto \iiint y^{2h_1+2h_2+2h_3-1}e^{iQ_1\bar{z}_{14}+iQ_2\bar{z}_{24}+iQ_3\bar{z}_{34}}z_{14}^{-2h_1}z_{24}^{-2h_2}z_{34}^{-2h_3} 
e^{\frac{iQ_1y^2}{z_{14}}+\frac{iQ_2y^2}{z_{24}}+\frac{iQ_3y^2}{z_{34}}}dydz_4d\bar{z}_4
\end{aligned}\end{equation}
We find that the integration of $y$ and $ \bar{z} $ can be performed independently. The $ \bar{z} $ integration gives
\begin{equation}
\int_{-\infty}^{\infty}e^{iQ_1\bar{z}_{14}+iQ_2\bar{z}_{24}+iQ_3\bar{z}_{34}}d\bar{z}_4=\delta_{Q_1+Q_2+Q_3,0}e^{\frac{1}{3}(Q_1-Q_2)\bar{z}_{12}+\frac{1}{3}(Q_2-Q_3)\bar{z}_{23}+\frac{1}{3}(Q_3-Q_1)\bar{z}_{31}}.
\end{equation}
The integration of $y$ is a little more subtle, but we can allow $y$ to have a small imaginary part and send the imaginary part to zero to obtain a convergent result. The contribution of $y$ integration is proportional to $ (\frac{iQ_1}{z_{14}}+\frac{iQ_2}{z_{24}}+\frac{iQ_3}{z_{34}})^{-h_1-h_2-h_3} $. 
Finally we have to integrate over $z_4$, and the result is proportional to
\begin{equation}
G\propto \int_{-\infty}^{\infty}z_{14}^{-2h_1}z_{24}^{-2h_2}z_{34}^{-2h_3}(\frac{iQ_1}{z_{14}}+\frac{iQ_2}{z_{24}}+\frac{iQ_3}{z_{34}})^{-h_1-h_2-h_3}dz_4
\end{equation}
The integral is difficult to perform, but we can make use of the charge neutrality condition, expressing one charge in terms of the other twos
\begin{equation}
G\propto \int_{-\infty}^{\infty}z_{14}^{-2h_1}z_{24}^{-2h_2}z_{34}^{-2h_3}(\frac{iQ_1z_{13}}{z_{14}z_{34}}+\frac{iQ_2z_{23}}{z_{24}z_{34}})^{-h_1-h_2-h_3}dz_4.
\end{equation}
Expanding the result, each term in the expansion has the form 
\begin{equation}\begin{aligned} 
&\int_{-\infty}^{\infty}z_{14}^{-2h_1}z_{24}^{-2h_2}z_{34}^{-2h_3}(\frac{z_{13}}{z_{14}z_{34}})^m(\frac{z_{23}}{z_{24}z_{34}})^{-h_1-h_2-h_3-m}dz_4 
\propto \frac{1}{z_{12}^{h_1+h_2-h_3}z_{23}^{h_2+h_3-h_1}z_{31}^{h_3+h_1-h_2}}. 
\end{aligned}\end{equation}
Combining the above results, we have the correlation function via  the Witten diagram 
\begin{equation}
G\propto \delta_{Q_1+Q_2+Q_3,0}\frac{e^{\frac{1}{3}(Q_1-Q_2)\bar{z}_{12}}e^{\frac{1}{3}(Q_2-Q_3)\bar{z}_{23}}e^{\frac{1}{3}(Q_3-Q_1)\bar{z}_{31}}}{z_{12}^{h_1+h_2-h_3}z_{23}^{h_2+h_3-h_1}z_{31}^{h_3+h_1-h_2}}, 
\end{equation}
which is the same as the three-point functions in a WCFT up to a constant. 

\par To summarize, by making an ansatz to respect the boundary conditions we obtain a bulk-boundary propagator that satisfies the bulk Klein-Gordon equation and matches the boundary correlator as the bulk point approaches the boundary. And we reconstruct  globally a  bulk field that reproduces the right form of the bulk-boundary propagator. By using the bulk field we obtain the bulk-bulk propagator which can be expressed in a bi-variate hypergeometric function. Using the bulk-boundary propagator we calculate the Witten diagrams for 3-point functions and find that it is in match with the one in WCFT. 

\section{ Local bulk reconstruction in AdS/WCFT} 

In this section, we study the local bulk reconstruction in AdS/WCFT. We will study the bulk proto-field in two different ways. One is to impose the bulk primary conditions to define the bulk proto-field.  This allows us to solve the proto-field and express it in terms of boundary quasi-primary operators.  The other one is to use the bulk-boundary OPE blocks, which allows us to compute the correlators of the bulk-boundary propagator with products of the stress tensors. 

\subsection{Algebraic definition of  bulk proto-field in AdS/WCFT} 

In AdS$_3$/CFT$_2 $, the conformal symmetry can be used to define an AdS proto-field as an exact linear combination of Virasoro descendants of a primary operator\cite{Anand:2017dav}. The AdS proto-field agrees with the gravitational perturbation theory when it is expanded   to all orders in the large $c$ limit.  It captures the information of gravitational background non-perturbatively. The requirement for the bulk proto-field is the bulk primary conditions
\begin{equation}L_{n}\Phi(y,0,0)|0\rangle=0,\hs{3ex}\bar{L}_{n}\Phi(y,0,0)|0\rangle=0,\hs{3ex}\forall n>2. \end{equation}
By requiring the bulk primary conditions and requiring the global part of the proto-field to match the global bulk field $\phi$, we can determine the AdS proto-field exactly. The bulk primary conditions come from the symmetry considerations. That is to consider the transformation of bulk proto-field under the action of Virasoro transformation. There is one unique extension of the boundary Virasoro transformations preserving the Fefferman-Graham(FG) gauge. And the resulting bulk Virasoro transformations that leave the bulk point invariant give rise to the bulk primary conditions. 

 The proto-field in the AdS/WCFT correspondence can be constructed in a similar way. Instead of the AdS$ _3 $ gravity with the Brown-Henneaux boundary condition which is dual to a 2D CFT, we consider the AdS$ _3 $ gravity with the CSS boundary condition. The CSS boundary condition is a Dirichlet-Neumann boundary condition. A generic solution with respect to the CSS boundary condition could be of the form \eqref{BanadosWAdS}. 
And it has been known that a generic Banados solutions can be related to the vacuum AdS metric through a diffeomorphism\cite{Roberts:2012aq}. For the metric (\ref{BanadosWAdS}),  we can  find a diffeomorphism to relate it to the vacuum AdS metric as well. Let the $ (w,\bar{w},u) $ coordinates be the vacuum AdS$ _3 $ coordinates
\be
 ds^2=\frac{du^2+dwd\bar{w}}{u^2}. 
 \ee
 The diffeomorphism is 
\begin{equation}\begin{aligned}
w&= f(z)-\frac{2y^2f'^2\bar{f}''}{4f'\bar{f}'+y^2f''\bar{f}''},\\
\bar{w}&= f({\bar{z}})-\frac{2y^2\bar{f}'^2f''}{4f'\bar{f}'+y^2f''\bar{f}''},\\
u&= y\frac{4(f\bar{f})^{\frac{3}{2}}}{4f'\bar{f}'+y^2f''\bar{f}'}. 
\end{aligned}\label{diffW}\end{equation}
  Here the functions $ f(z) $ and $ f({\bar{z}}) $ satisfy the equations
  \be
   S(f,z)=\frac{12}{c}L(z), \hs{3ex}  f(\bar{z})= e^{\sqrt{-\frac{24}{c}\bar{L}}(\bar{z}-B(z))}, \label{diffW2}\ee 
   where $S(f,z)$ stands for the Schwarzian derivative 
   \be 
   S(f,z)=\frac{f'''f'-\frac{3}{2}{f''}^2}{(f')^2}.\ee
    The translation along $ \bar{z} $  leaves $ \bar{L} $ invariant, and the exponential map will change $\bar{L}$ by a constant. The exponential map is needed to relate AdS$ _3 $ with the CSS boundary conditions to the  AdS vacuum. However there are ambiguities in the exponential map. The exponential map can differ by an arbitrary constant times, but this will not change the discussion below. 
    
We then look for the bulk generalizations of boundary Virasoro-Kac-Moody transformations. The bulk Virasoro-Kac Moody transformations need to satisfy two requirements, 
\begin{enumerate}
\item They must agree with the Virasoro-Kac-Moody transformations on the boundary,
\item They preserve the metric form in the FG gauge.
\end{enumerate}
 We suppose that the bulk Virasoro-Kac-Moody transformation is of the form
\begin{equation}(y,z,\bar{z},L,B)\to(\tilde{y},\tilde{z},\tilde{\bar{z}},\tilde{L},\tilde{B})\end{equation}
And we use $F$ to denote the transformations (\ref{diffW}) obtained above. To preserve the metric form in the FG gauge, we can map an asymptotic AdS space characterized by $(L,B)$  to the AdS vacuum and then map the AdS vacuum to another asymptotic AdS characterized by $(\tilde{L},\tilde{B})$. The bulk Virasoro-Kac-Moody transformations have to satify
\begin{equation}F(y,z,\bar{z},L,B)=\tilde{F}(\tilde{y},\tilde{z},\tilde{\bar{z}},\tilde{L},\tilde{B}).\end{equation}
Also it has to agree with the infinitesimal boundary Virasoro-Kac-Moody transformations 
\be
 \epsilon L_m(y,z,\bar{z})|_{y\to 0}=\epsilon(0,z^{m+1},0),\hs{3ex} \epsilon P_n (y,z,\bar{z})|_{y\to 0}=\epsilon(0,0,z^n), \ee 
where $ L_m $ is the infinitesimal Virasoro transformation 
\be
L_m(y,z,\bar{z})=(\d_my,\d_m z, \d_m\bar{z}),
\ee
and $ P_n $ is the infinitesimal Kac-Moody transformation
\be
P_n(y, z,\bar{z})=(\d_ny,\d_n z, \d_n\bar{z}).
\ee
The solution for the bulk Virasoro-Kac-Moody transformations is
\bea
\delta_{m}y&=&\frac{1}{2}(m+1)yz^{m},\\ 
\delta_{m}z&=&\frac{z^{m-1}((m^2+m+z^2S(z))\bar{S}(\bar{z})y^4-4z^2)}{y^4S(z)\bar{S}(\bar{z})-4},\\
\delta_{m}\bar{z}&=&\frac{2m(m+1)y^2z^{m-1}}{y^4S(z)\bar{S}(\bar{z})-4},
\eea
\begin{equation}\delta_{n}y=0,\hs{3ex}\delta_{n}z=0,\hs{3ex}\delta_{n}\bar{z}=z^n.\end{equation}
where the anti-holomorphic Schwarzian is a constant. We can check that the bulk Virasoro-Kac-Moody transformations agree with Virasoro-Kac-Moody transformation on the boundary. Moreover the bulk Virasoro-Kac-Moody transformations preserve the form of the AdS metric with the CSS boundary conditions and keeping  the anti-holomorphic $\bar{L}$ unchanged.

The bulk scalar is invariant under coordinate transformation $\Phi(y,z,\bar{z})=\Phi(y',z',\bar{z}')$. The Viraroso-Kac-Moody transformation acts on a scalar as 
\bea
L_m\Phi(y,z,\bar{z})&=&(\delta_my\partial_y+\delta_mz\partial_z+\delta_m\bar{z}\partial_{\bar{z}})\Phi(y,z,\bar{z}), \nn\\
P_n\Phi(y,z,\bar{z})&=&(\delta_ny\partial_y+\delta_nz\partial_z+\delta_n\bar{z}\partial_{\bar{z}})\Phi(y,z,\bar{z}). 
\eea
 It is easy to see the  point $(y,0,0)$ is invariant under $ L_m $ for $ m\geq 2 $ and $ P_n $ for $ n\geq 1 $. Using the transformation law on a scalar,  we find the bulk primary condition
 \begin{equation}L_{m}\Phi(y,0,0)|0\rangle=0, \hs{3ex}m\geq 2, \label{bulkprimary1}\end{equation}
\begin{equation}P_{n}\Phi(y,0,0)|0\rangle=0 \hs{3ex}n\geq 1. \label{bulkprimary2}  \end{equation}
 It is sufficient to determine the bulk proto-field. The bulk Kac-Moody primary condition is easy to solve. 
 And the bulk Virasoro  primary condition can be used to solve the bulk proto-field exactly. The solution can be written in an expansion of  boundary operator 
\begin{equation}\Phi(y,0,0)|0\rangle=\sum_{N=0}^{\infty}y^{2h+2N}|\Phi_{N}\rangle \end{equation}
where $| \Phi_N\rangle $ is a level $N$ descendant of $\mO$ in the holomorphic sector, as the anti-holomorphic sector is trivial. The conditions \eqref{bulkprimary1}\eqref{bulkprimary2} on $\Phi(y,0,0)$ is equivalent to 
\bea
L_m | \Phi_N\rangle&=&0,  \hs{3ex}m\geq 2 \nn\\
P_n | \Phi_N\rangle&=&0.  \hs{3ex}n\geq 1. 
\eea
 We can solve it in terms of quasi-primaries and their global descendants. The result is
\begin{equation}
|\Phi_{N}\rangle=\lambda_{N}\left(\frac{L_{-1}^N}{{|L_{-1}^N}\mO|^2}+\frac{L_{-N}^{quasi}}{|L_{-N}^{quasi}\mO|^2}+\frac{L_{-1}L_{-N+1}^{quasi}}{|L_{-1}L_{-N+1}^{quasi}\mO|^2}+...\right)|\mO\rangle \end{equation}
where $\mO$ is the boundary primary and the coefficient is determined such that the global part (the first term) agrees with the global reconstruction, and reproduces the right global bulk-boundary propagator. We have already obtained the bulk-boundary propagator of AdS/WCFT, so we can use it to determine the coefficient
\begin{equation}\lambda_{N}=(-iQ)^N. \end{equation}
Then for the global part of $\Phi$ we recover the previous expression for the bulk field. 

At low levels, we can solve the proto-field explicitly. Obviously $|\Phi_0\rangle = |\mO\rangle$, and 
\be
|\Phi_1\rangle=-iQ \frac{1}{2h}L_{-1}|\mO\rangle, 
\ee
and 
\be
|\Phi_2\rangle=-Q^2\frac{(2h+1)(c+8h)}{(2h+1)c+2h(8h-5)}\left(L^2_{-1}-\frac{12h}{c+8h}L_{-2}\right)|\mO\rangle, 
\ee
where $h$ is the conformal weight of $\mO$. 
One remarkable fact is that in the large $c$ limit, the terms $L^N_{-1}$ dominates in $\Phi_N$, suggesting that the theory is effectively free and only the global blocks survive. 

The bulk proto-field is a field that sees the gravitational background of the AdS physics to all orders. We can obtain a recursion relation for the correlators involving the stress tensor, the bulk proto-field and the boundary primary as in \cite{Anand:2017dav} . The recursion relations share  the same structure as in \cite{Anand:2017dav}. 

\par With the  reconstructed proto-field and the boundary primary correlator, we can read the bulk-bulk  propagator of two proto-scalar fields directly. Using the fact that the correlator of distinct quasi-primaries vanish, we can single out the contributions from a certain quasi-primary. The contribution to $ \Phi $ from a level-$n$ quasi-primary is
\begin{equation}
|\Phi^{(n)}\rangle=y^{2h+2n}\sum_{m=0}^{\infty}y^{2m}\lambda_{n+m}\frac{L_{-1}^mL_{-n}^{quasi}|\mO\rangle}{|L_{-1}^mL_{-n}^{quasi}\mO|^2}. \end{equation}
A useful identity for computing the propagator is
\begin{equation}|L_{-1}^mL_{-n}^{quasi}\mO|^2=(2h+2n)_mm!|L_{-n}^{quasi}\mO|^2 \end{equation}
The correlator of two quasi-primaries is
\begin{equation}\langle L_{-n}^{quasi}\mO_1L_{-n}^{quasi}\mO_2\rangle=\frac{(-1)^ne^{iQ\bar{z}_{12}}|L_{-n}^{quasi}\mO|^2 }{z_{12}^{2h+2n}}\end{equation}
So the contribution to the propagator of the proto-field from a level-$n$ quasi-primary is
\bea
\lefteqn{\langle \Phi(y_1,z_1,\bar{z}_1)^{(n)}\Phi(y_2,z_2,\bar{z}_2)^{(n)}\rangle}\nn\\
&=&\frac{(y_1y_2)^{2h+2n}e^{iQ\bar{z}_{12}}}{|L_{-n}^{quasi}\mO|^2}\sum_{m,m'=0}^{\infty}\frac{(iQ)^{2n}(-iQy_1^2)^m(-iQy_2^2)^{m'}}{m!m'!(2h_n)_m(2h_n)_{m’}}
 {\partial_{z_1}}^{m}{\partial_{z_2}}^{m'}(\frac{1}{z_{12}^{2h+2n}})\nn\\
 &=&\frac{(\frac{y_1y_2}{z_{12}^2})^{2h+2n}(iQ)^{2n}e^{iQ\bar{z}_{12}}}{|L_{-n}^{quasi}\mO|^2}\sum_{m,m'=0}^{\infty}\frac{(\frac{iQy_1^2}{z_{12}})^m(\frac{iQy_2^2}{z_{12}})^{m'}}{m!m'!}\frac{(2h_n)_{m+m'}}{(2h_n)_m(2h_n)_{m’}}. \label{quasicorrelator}
\eea
When $n=0$, we get the global bulk-bulk scalar propagator, which is exactly what we got in the previous section. There is one interesting property of the result. The correlator of $n$-level quasi-primary is proportional to global bulk-bulk propagator with a different scaling dimension $2h+2n$. 

The expression \eqref{quasicorrelator} is a bi-variate hypergeometric function that does not have a closed form. Recall that in the AdS/CFT correspondence, the bulk-bulk propagator have a simple dependence on the geodesic distance. But in the warped case the global symmetry is no longer manifest, and the geodesic distance is no longer an invariant, so one could expect  that the bulk scalar propagator has no such simplification.

\subsection{The Bulk-boundary OPE blocks in AdS/WCFT} 

 In a conformal field theory, the product of the primary operators can be expanded in term of the primary modules. In 2D CFT the Virasoro block can be computed in a recursive way. In particular, in the large central charge limit, the vacuum block dominates\cite{Hartman:2013mia,Chen:2013kpa}.  For a warped CFT, it is expected that the vacuum module dominates in the large $c$ limit as well\cite{Fitzpatrick:2015zha,Apolo:2018eky}.  

Another way of studying the bulk proto-field is by using the bulk-boundary OPE blocks. The basic idea is to take the bulk proto-field $\Phi(y,z,\bar{z})$ as an operator in WCFT, and consider the OPE blocks involving $\Phi(y,z,\bar{z})$ and other primary operators. In particular, as we are interested in probing the nonperturbative information of the gravity, we focus on the scalar  OPE block involving the vacuum module 
\be
\Phi(y,0,0)\mO(z,\bar{z})=y^{2h}z^{-2h}e^{iQ\frac{z\bar{z}+y^2}{z}}+\cdots.  
\ee
The first term is simply the bulk-boundary propagator, coming from the identity operator.  And the ellipsis consists of the contribution from the descendent operators in the vacuum module, it encodes the information of the correlators 
\be
\langle \Phi(y,0,0)\mO(z,\bar{z})T(z_1) \cdots T(z_n) \rangle. 
\ee

The bulk-boundary OPE block  can be calculated by applying a diffeomorphism to the bulk-boundary propagator for pure AdS. In the last section, we have discussed the diffeomorphism between the pure AdS vacuum and the AdS solution \eqref{BanadosWAdS} with nonvanishing stress tensor $T(z)$ and $U(1)$ current $B(z)$. The diffeomorphism is given by the relations \eqref{diffW}\eqref{diffW2}. Now we take this diffeomorphism as an operator equation, and obtain an operator-valued AdS metric which depends on the operator $T(z)$ and $B(z)$. After applying this diffeomorphism to the bulk-boundary propagator, we obtain the vacuum sector of the bulk-boundary OPE block
\begin{equation} \Phi(y,z_2,\bar{z}_2)O(z_1,\bar{z}_1)=(w'(z_1)w'(\bar{z}_1))^h(\frac{u}{w_{12}})^{2h}e^{iQ(\bar{w}_{12}+\frac{y^2}{w_{12}})} , \label{OPEblock}
\end{equation}
where $ (u,w,\bar{w}) $ are the pure AdS coordinates. To obtain the diffeomorphism, we have to solve the holomorphic function satisfying $ S(f,z)=\frac{12}{c}T(z) $. We solve the equation order by order in $1/c$,
\begin{equation} 
f(z)=z+\frac{1}{c}f_1(z)+\frac{1}{c^2}f_2(z)+.... 
\end{equation}
The first two functions $f_1$ and $f_2$ satisfy the differential equations 
\begin{equation} \begin{aligned}
&f_1'''(z)-12T(z)=0, \\
&f_1'''(z)f_1'(z)+3f_1''(z)^2-2f_2'''(z)=0. 
\end{aligned} \end{equation}
By integration, we get 
\begin{equation}\begin{aligned}
&f_1(z)=6\int_{0}^{z}dz'(z-z')^2T(z'), \\
&f_2(z)=72\int_{0}^{z}dz'\int_{0}^{z'}dz''(z-z')^2(z-z'')T(z')T(z''). 
\end{aligned}\end{equation}
Then we can expand \eqref{OPEblock} and get the bulk-boundary OPE block
\begin{equation}\begin{aligned}
\log \Phi(y,0,0)O(z,\bar{z})=&\log(\frac{y^{2h}}{z^{2h}}e^{iQ(\bar{z}+\frac{y^2}{z})})+\frac{hf_1'}{c}-\frac{2hf_1}{cz}-iQ\frac{y^2f_1}{cz^2}\\
&+\frac{h(2f_2'-f_1'^2)}{2c^2}-\frac{2hf_2}{c^2z}+\frac{hf_1^2}{c^2z^2}-iQ\frac{y^2f_2}{c^2z^2}+iQ\frac{y^2f_1^2}{c^2z^3}+\cdots.
\end{aligned}\end{equation}
The bulk-boundary OPE blocks have no dependence on the anti-holomorphic coordinate since only translations are allowed for $ \bar{z} $.
The order $1/c$ terms can be combined into the kernel 
\begin{equation}
K_T=\frac{6}{c}\int_0^zdz'\left(2h\frac{z'(z-z')}{z}+iQ\frac{y^2(z-z')^2}{z^2}\right)T(z'). 
\end{equation}
In the $y \to 0 $ limit, it reproduce the kernels for  ``boundary-boundary" $\mO(z)\mO(0)$ OPE block\cite{Fitzpatrick:2016mtp}. 
At the next order we obtain the kernel 
\begin{equation}
K_{TT}=\frac{72}{c^2}\int_0^zdz'\int_0^{z'}dz''\left(\frac{h(z-z')^2z''^2}{z^2}-iQ\frac{y^2(z-z')^2(z-z'')z''}{z^3}\right)T(z')T(z''). 
\end{equation}
Different from the usual AdS/CFT case, here  are only the kernels involving the holomorphic stress tensors. 

Using the bulk-boundary OPE block, we can calculate the correlators of the bulk-boundary propagator with products of local stress tensors. The simplest one is $ <\Phi \mO T> $. It can be obtained using only the $ K_T $ kernel
\begin{equation}\frac{\langle \Phi(y,0,0)\mO(z,\bar{z})T(z_1)\rangle}{\Phi(y,0,0)\mO(z,\bar{z})}=\langle K_TT(z_1)\rangle=\frac{hz^2}{z_1^2(z_1-z)^2}+iQ\frac{y^2z}{z_1^3(z_1-z)}.
\end{equation}
From the correlators of the bulk-boundary propagator with one stress tensor, we can read the coefficients of Virasoro descendants in the expansion of the bulk proto-field. We check that the coefficient in the lowest order agrees with the algebraic calculation. For example, let us consider the level-two quasi-primary operators. The only level-two quasi-primary operator is
\begin{equation}
L_{-2}^{quasi}=L_{-1}^2-\frac{12h}{c}L_{-2}.
\end{equation}
The quasi-primary operator contribution to the bulk proto-field is
\begin{equation}
\delta\Phi_2\approx \frac{(-iQ)^2y^4}{2(2h)_2}(L_{-1}^2-\frac{12h}{c}L_{-2})\mO(0)
\end{equation}
We have obtained the correlator of bulk boundary OPE with one stress tensor. We subtract the contribution of the global bulk field. Expanding in $y$ coordinate, the first order is
\begin{equation}
\langle (\Phi(y,0,0)-\Phi^{global}(y,0,0))\mO(z,\bar{z})T(z_1)\rangle=-\frac{3h(iQ)^2y^4e^{iQ\bar{z}}}{(2h)_2z_1^4z^{2h}}+...
\end{equation}
The level two quasi-primary operator have the following stress tensor correlator
\begin{equation}
\langle (L_{-2}\mO(0))O(z,\bar{z})T(z_1)\rangle=\frac{ce^{iQ\bar{z}}}{2z^{2h}z_1^4}, 
\end{equation}
so we can read the coefficient of the quasi-primary in the proto-field
\begin{equation}
\delta\Phi_2\approx -\frac{(-iQ)^2y^4}{2(2h)_2}\frac{12h}{c}L_{-2}\mO(0). 
\end{equation}
This agrees with the result using  algebraic method.


\section{Conclusion}
The AdS/WCFT correspondence serves as one  example of holographic duality beyond AdS/CFT. In this paper, we try to understand how to reconstruct the AdS field in terms of WCFT operators. By imposing a specific boundary condition, we get a vacuum bulk-boundary propagator that reproduces the WCFT correlator as the bulk point approaches boundary. And we use it to globally reconstruct the bulk scalar and calculate the vacuum bulk-bulk propagator. Furthermore we  reconstruct locally a bulk proto-field which knows the gravitational background non-perturbatively. We studied the bulk proto-field using two different ways. One is to impose the bulk primary condition on the proto-field, while the other is to use the bulk-boundary OPE block. The two ways lead to consistent picture. By solving the bulk primary condition, the bulk proto-field  can be solved exactly. 

\par In 2D CFT,  the OPE blocks can be calculated using the $ SL(2,R)\times SL(2,R) $ Wilson lines\cite{Fitzpatrick:2016mtp}. And the bulk-boundary OPE blocks can be calculated using bulk $ SL(2,R)\times SL(2,R) $ Wilson lines. The Wilson line calculation is equivalent to the calculation using  the AdS diffeomorphism\cite{Anand:2017dav}. It would be interesting to investigate if the Wilson line method still works in AdS/WCFT or in WCFT. 

\section*{Acknowledgments}

 We would like to thank Peng-xiang Hao, Apolo Luis and Wei Song for valuable discussions.  The work is supported in part by NSFC Grant No. 11275010, No. 11325522, No.~11335012
and No. 11735001.

\bibliographystyle{unsrt}

\end{document}